\documentclass[graybox]{svmult}

\usepackage{mathptmx}       
\usepackage{helvet}         
\usepackage{courier}        
\usepackage{type1cm}        
\usepackage{makeidx}         
\usepackage{graphicx}        
\usepackage{multicol}        
\usepackage[bottom]{footmisc}

\usepackage{amsmath}

\newcommand{\tmpnote}[1]%
   {\begingroup{\color{blue}\it (FIXME: #1)}\endgroup}

\makeindex             

\begin{document}

\title*{Flat bands as a route to high-temperature superconductivity in graphite}
\author{T.T. Heikkil\"a and G.E. Volovik}
\institute{T.T. Heikkil\"a \at University of Jyvaskyla, Department of
  Physics and Nanoscience Center, P.O. Box 35, FI-40014 University of
  Jyv\"askyl\"a, Finland \email{Tero.T.Heikkila@jyu.fi}
\and G.E. Volovik \at Aalto University, Department of Applied Physics
and Low Temperature Laboratory, P.O. Box 15100, FI-00076 AALTO,
Finland and L.~D.~Landau Institute for
Theoretical Physics, 117940 Moscow, Russia \email{volovik@boojum.hut.fi}}%
%

\maketitle

\abstract{Superconductivity is traditionally viewed as a
  low-temperature phenomenon. Within the BCS theory this is understood
  to result from the fact that the pairing of electrons takes place
  only close to the usually two-dimensional Fermi surface residing at
  a finite chemical potential. Because of this, the critical
  temperature is exponentially suppressed compared to the microscopic
  energy scales. On the other hand, pairing electrons around a
  dispersionless (flat) energy band leads to very strong
  superconductivity, with a mean-field critical temperature linearly proportional to the microscopic coupling constant. The prize to be paid is that flat bands can generally be generated only on surfaces and interfaces, where high-temperature superconductivity would show up. The flat-band character and the low dimensionality also mean that despite the high critical temperature such a superconducting state would be subject to strong fluctuations. Here we discuss the topological and non-topological flat bands discussed in different systems, and show that graphite is a good candidate for showing high-temperature flat-band interface superconductivity.
 }

The purpose of this chapter is to propose a route to increasing the
critical temperature of superconductivity by searching for special
electronic dispersion that would promote the superconducting
strength. We first show that a huge increase in the (mean-field) critical
temperature is possible, if a dispersionless energy spectrum, a {\it
  flat band} can be created in the system in the absence of the
interaction leading to superconducting correlations. We then discuss a
few known schemes to generate such (approximate or exact) flat bands.

\section{Superconductivity: pairing energy vs. dispersion}
\label{pairing}


Within the BCS mean-field theory, the occurrence of Cooper pairing at zero temperature can be studied via the free energy density for the pairing energy $\Delta$: 
\begin{equation}
F_\Delta = -\frac{1}{2} \int \frac{d^d p}{(2\pi \hbar)^d} \left(E_{\bf p}(\Delta)-E_{\bf p}(\Delta=0)\right)+ \frac{\Delta^2}{2|g|},
\label{eq:freeenergy}
\end{equation}
where $d$ is the dimensionality, $g<0$ describes the interaction strength, and $E_{\bf p}(\Delta)=\sqrt{\epsilon_{\bf p}^2+\Delta^2}$ is the quasiparticle excitation energy at momentum value ${\bf p}$, evaluated in a system with the normal-state dispersion $\epsilon_{\bf p}$. The first term in Eq.~\eqref{eq:freeenergy} demonstrates that the formation of a gap $\Delta$ decreases the energy of quasiparticles, which fill the negative energy levels of Dirac vacuum.  The second term is the cost of the formation of the gap, which perturbs the vacuum. For simplicity, we consider  spinless fermions and the gap that does not depend on momentum.

Requiring $\Delta$ to minimize $F_\Delta$, we get the self-consistency relation
\begin{equation}
\Delta = \frac{|g|}{2}  \int \frac{d^d p}{(2\pi \hbar)^d} \frac{\Delta}{E_{\bf p}(\Delta)}\,,
\label{selfconsistent1}
\end{equation}
or
\begin{equation}
1 = \frac{|g|}{2}  \int \frac{d^d p}{(2\pi \hbar)^d} \frac{1}{E_{\bf p}(\Delta)}\,.
\label{selfconsistent2}
\end{equation}
Equation (\ref{selfconsistent2})  dictates the behavior of $\Delta$ at different dimensionalities $d$ and for different normal-state energy spectra $\epsilon_{\bf p}$. 

For $s$-wave superconductivity in conventional metals with an
isotropic Fermi surface and dispersion $\epsilon=v_F(p-p_F)$ expanded around
the Fermi energy $\epsilon_F=v_F p_F$, the integral in Eq.~(\ref{selfconsistent2}) is concentrated in the vicinity of the Fermi surface
\begin{equation}
1 = \frac{|g| A_d}{(2\pi \hbar)^d v_F} \int_0^{\epsilon_{\rm uv}}
\frac{d\epsilon}{\sqrt{\epsilon^2+\Delta^2}} \overset{\epsilon_{\rm
    uv} \gg \Delta}{\approx} |g| \nu_F \ln \frac{\epsilon_{\rm uv}}{\Delta}\,.
\label{logarithm}
\end{equation}
Here $\nu_F=\frac{A_d}{(2\pi \hbar)^d v_F}$ is the density of states
in the normal metal; $A_d$ is the area of the $d$-dimensional Fermi
surface; and $\epsilon_{\rm uv}\ll \epsilon_F$ is the ultraviolet cut-off of the logarithmically diverging integral, such as the Debye temperature.
This leads to the exponentially suppressed gap: 
\begin{equation}
\Delta = \epsilon_{\rm uv}  \exp\left(- \frac{1}{ |g| \nu_F}\right) \,,
\label{exponent}
\end{equation}
and correspondingly to the exponential suppression of the transition temperature $T_c$. 

Situation drastically changes when the spectrum of the normal state has a flat band -- a region in momentum ${\bf p}$ where $\epsilon_{\bf p}=0$.
Since within the flat band $E_{\bf p}(\Delta)=\Delta$, Eq.~(\ref{selfconsistent2}) becomes
\begin{equation}
1 = \frac{|g| V_d}{2(2\pi \hbar)^d \Delta}\,,
\label{nonlogarithm}
\end{equation}
where $V_d$ is the volume of the flat band in momentum space.
Hence  instead of the usual exponentially suppressed behavior in Eq.~(\ref{exponent}) we have the gap
 $\Delta$ that is linearly proportional to the interaction strength:
\begin{equation}
 \Delta = \frac{|g| V_d}{2(2\pi \hbar)^d}\,,
\label{linear}
\end{equation}
Since the critical temperature of the superconductors is typically of the same order of magnitude as the gap at zero temperature, the resulting superconductivity may exist at high temperatures.

It is instructive to consider the intermediate case when the quasiparticle spectrum is
\begin{equation}
\epsilon(p) = \epsilon_0\left(\frac{p}{p_0}\right)^M \,.
\label{powerM}
\end{equation}
Then for $M>d$, Eq.~(\ref{selfconsistent2}) gives  the power law dependence of transition temperature
on the coupling constant: 
\begin{equation}
\Delta \propto |g|^\frac{M}{M-d} \,.
\label{powerMd}
\end{equation}
 In the limit of large $M\gg 1$, the spectrum (\ref{powerM})
 transforms to the flat band concentrated at $p<p_0$, and the gap
 (\ref{powerMd}) asymptotically approaches  the linear dependence on
 the coupling $g$ in Eq.~(\ref{linear}). The case with $d=1$ and
 $M=2$, where $\Delta \propto g^2$, has been considered by Kopaev
 \cite{Kopaev1970} and
 Kopaev-Rusinov \cite{KopaevRusinov1987}.

In the sections below, we consider different systems where the exact or approximate flat bands could be realized.

\section{Flat band induced by interaction}
\subsection{Landau phenomenology}

We start with the flat bands induced by interaction between the fermions. As was found by 
Khodel and Shaginyan \cite{Khodel1990}, the interaction may lead to the merging of different fermionic energy levels, which results in the formation of a dispersionless band, see also 
Refs.  \cite{Volovik1991,Nozieres92}.
The effect of the merging of discrete energy levels due to interaction
has been reported in a recent paper, Ref. \cite{Dolgopolov2014}.

 \begin{figure}
 \begin{center}
  \includegraphics[width=0.8\linewidth]{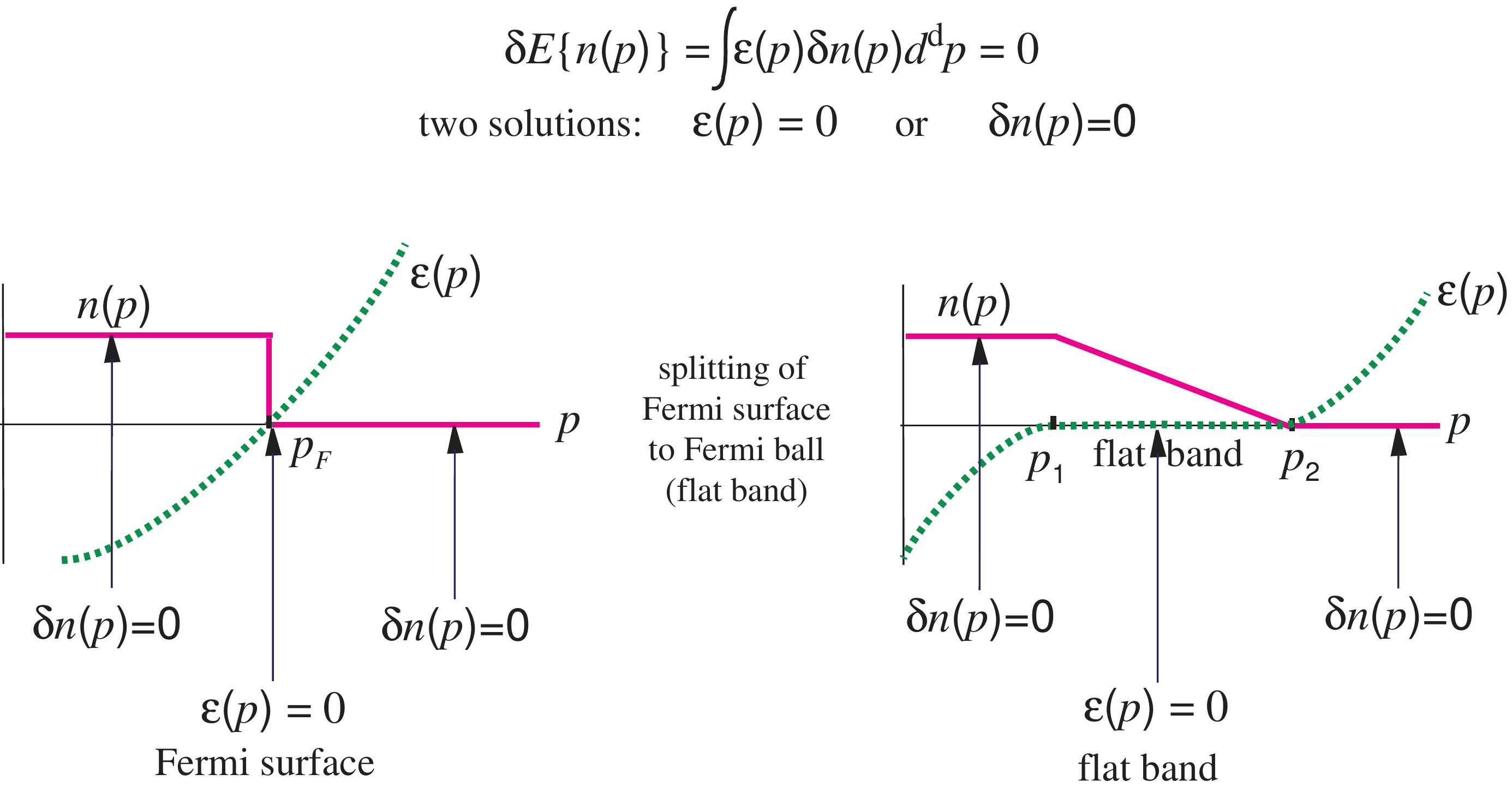}
\end{center}
  \caption{\label{KhodelFigure}
Illustration of the formation of the Khodel-Shaginyan flat band due to
interaction between fermions.
({\it Top}): the Landau model of the Fermi liquid considers the energy  ${\cal E}\{n({\bf p})\}$ as the functional  
of the distribution function $n({\bf p})$ of quasiparticles. The variation of the functional gives two types of solutions: 
$\epsilon_{\bf p}=0$ or 
$\delta n_{\bf p}=0$. ({\it Left}): distribution of quasiparticles in the class of Fermi liquids. Two regions 
 with solutions $\delta n_{\bf p}=0$ are separated by the surface with
 solution $\epsilon_{\bf p}=0$. This is the Fermi surface. Topological
 stability of this class is demonstrated in
 Fig. \ref{FermiSurfaceQPTFig}.  ({\it Right}): For strong interaction
 (large interaction constant), the intermediate region with the solution $\epsilon_{\bf p}=0$, may become finite. This is the flat band.  Topological structure of such flat band is demonstrated in Fig. 
 \ref{KhodelTopologyFig}.
 }
\end{figure}

Their argument is based on the phenomenological Landau's consideration on the derivation of the distribution function $n_{\bf p}$ of fermions  at $T=0$. It is determined by the energy functional ${\cal E}\{n({\bf p})\}$, whose  variational derivative $\epsilon_{\bf p}$ is the quasiparticle energy. The variation of this functional gives the equation for $n_{\bf p}$ and $\epsilon_{\bf p}$:
\begin{equation}
\delta {\cal E}\{n({\bf p})\} = \int \frac{d^d p}{(2\pi \hbar)^d}~ \epsilon_{\bf p} \delta n_{\bf p} =0
\,.
\label{variation}
\end{equation}
Since the quasiparticle distribution function is constrained by the
Pauli principle $0\leq n_{\bf p}\leq 1$, there are two classes of
solutions of the variational problem. One class is $\epsilon_{\bf
  p}=0$, which is valid if  $0< n_{\bf p} <1$; another one is $\delta
n_{\bf p}=0$ with  $n_{\bf p} =0$ or $n_{\bf p} =1$. In the following
we illustrate these two classes of solutions.

\subsection{Landau Fermi liquid and its topological stability}

Let us start with  the Fermi gas -- the system of free fermions with the spectrum $\epsilon_{\bf p}= p^2/2m - \mu$, where $\mu >0$. The energy functional for free fermions is ${\cal E}\{n({\bf p})\} = \int \frac{d^d p}{(2\pi \hbar)^d} ~\epsilon_{\bf p}n_{\bf p}$, which gives the solution shown in Fig. \ref{KhodelFigure} ({\it left}). The solution of the class $\epsilon_{\bf p}=0$ forms the Fermi surface with radius $p_F$, where $p_F^2/2m = \mu$. Outside of the Fermi surface
the distribution function $n_{\bf p}={\rm const}$, with $n_{\bf p}=1$ for $p<p_F$ and $n_{\bf p}=0$ for $p>p_F$. This corresponds to the class of solutions with $\delta n_{\bf p}=0$.

The structure with the Fermi surface is topologically protected. The topological stability can be illustrated using the $d=2$ system, when the Fermi surface $p=p_F$ is the 1d circle. For that one has to consider the Green's function at imaginary frequency:
\begin{equation}
G(\omega,{\bf p})= \frac{1}{i\omega - \left(  \frac{p^2}{2m} - \mu \right)} \,.
\label{GreenFunction}
\end{equation}
\begin{figure}[t]
\centerline{\includegraphics[width=0.3\linewidth]{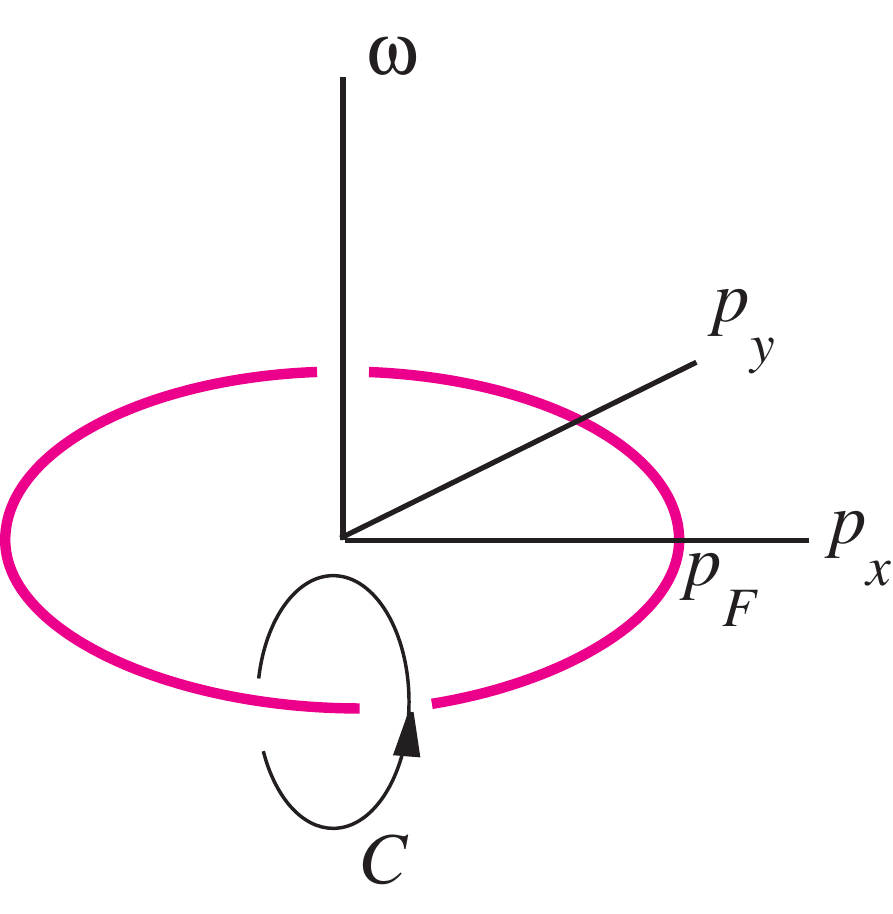}}
\medskip
\caption{Illustration of the topological stability of the Fermi
  surface on an example case with dimension $d=2$, when the Fermi
  surface forms a closed loop. Green's function has singularities on
  the line $\omega=0$, $p_x^2+p_y^2=p_F^2$ in the three-dimensional space $(\omega,p_x,p_y)$.
 Stability of the Fermi surface is protected by the invariant (\ref{InvariantForFS}) which is represented by an integral  over an arbitrary contour $C$ around the Green's function singularity.
}
\label{FermiSurfaceQPTFig}
\end{figure}

The Green's function in Eq.~(\ref{GreenFunction}) has singularities  at $\omega=0$ for  ${\bf p}$ belonging to the Fermi surface. 
These points form a closed line in the three dimensional $(\omega,p_x,p_y)$-space, see 
Fig. \ref{FermiSurfaceQPTFig} for $d=2$. 
This line has a topological winding number: the phase $\Phi$ of the Green's function, $G=|G|e^{\Phi}$
changes by $2\pi$ along an arbitrary contour $C$ around the line. In other words, the Fermi surface represents the  $p$-space  analog of the vortex lines in superfluids and superconductors, where the phase of the order parameter changes by $2\pi$ around the vortex. The $2\pi$ winding of the phase 
$\Phi$ cannot change under small deformations of the parameters of the system, and thus is robust to the interactions between the particles, if we do not consider  the superconducting, magnetic or  other phase transition, which drastically (non-perturbatively) reconstructs the energy spectrum. This topological stability is the reason why interacting Fermi liquids preserve the Fermi surface.

For more complicated cases, when the Green's function has spin, band and other indices, and for arbitrary dimension $d$
the  winding number
$N$ of the Fermi surface is expressed analytically  in terms of the matrix Green's function in the following form:
\begin{equation}
N={\bf tr}~\oint_C \frac{dl}{2\pi i}  G(\omega,{\bf p})\partial_l
G^{-1}(\omega,{\bf p})\,.
\label{InvariantForFS}
\end{equation}
Here the integral is taken over an arbitrary contour $C$ around the Green's function singularity
in  the $d+1$
momentum-frequency space. For the Green's function in Eq.~(\ref{GreenFunction}) the topological invariant $N=1$.

\subsection{Khodel-Shaginyan flat band and its topology}

When the interaction between the particles is strong enough, so that it starts dominating over the 
fermionic statistics, a more classical behavior of the distribution
function may emerge, in which the natural solution of the variational
problem corresponds to a zero value of the variational derivative, 
$\delta {\cal E}\{n({\bf p})\}/\delta n_{\bf p}=0$.  Therefore, with an increasing interaction strength one may expect the
topological quantum phase transition to the  distribution in
Fig. \ref{KhodelFigure} ({\it right}), where the solution
$\epsilon_{\bf p}=0$ is spread over a finite region in momentum space,
i.e., forming a flat band.

\begin{figure}[t]
\centerline{\includegraphics[width=0.3\linewidth]{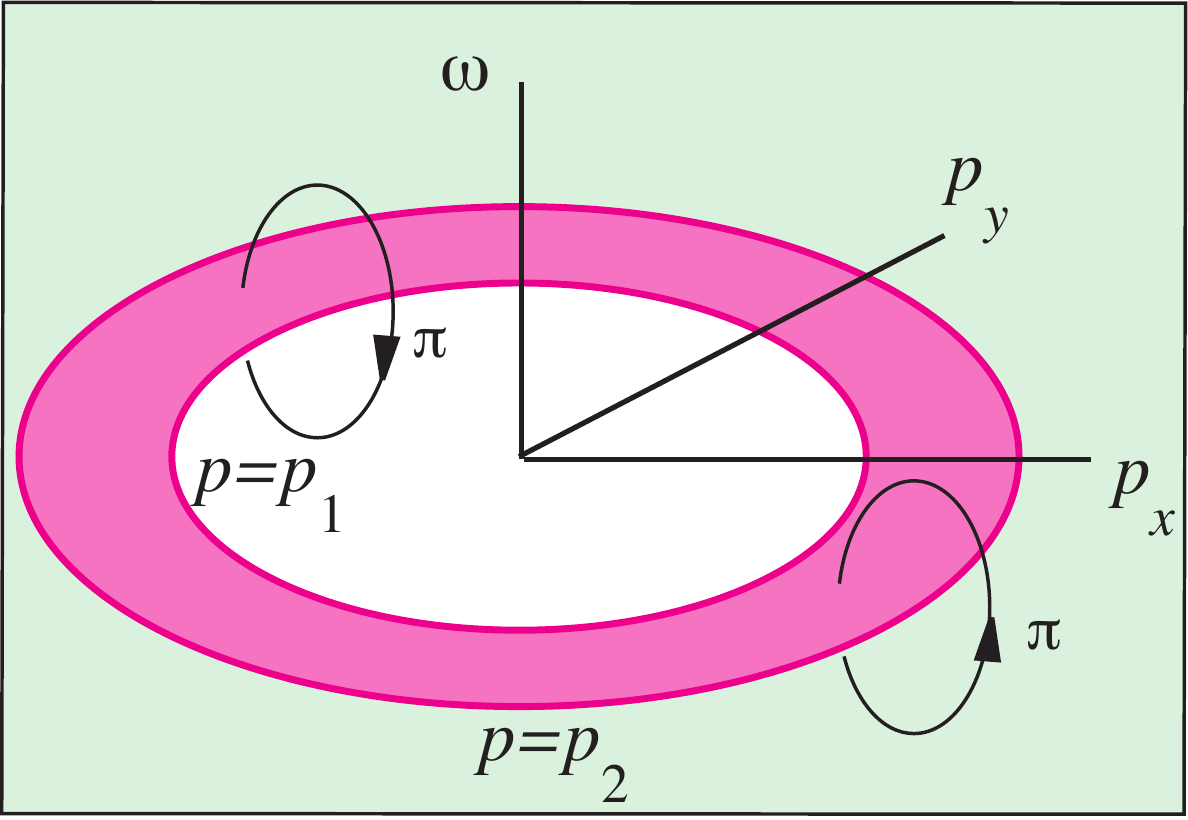}}
\caption{Illustration of the topological structure of the Khodel-Shaginyan flat band in a $d=2$ system.
According to Fig.~\ref{KhodelFigure} ({\it right}) the Fermi surface at $p=p_F$ spreads into a flat band concentrated in the region $p_1<p<p_2$. Correspondingly the line of the Green's function singularities in Fig. \ref{FermiSurfaceQPTFig} -- a vortex line -- is spreading to an analog of a domain wall terminating on a pair of $\pi$-vortices at $p=p_1$ and $p=p_2$. \cite{Volovik1991}
}
\label{KhodelTopologyFig}
\end{figure}

 The topological structure of the Khodel-Shaginyan flat band is shown in Fig. 
 \ref{KhodelTopologyFig}. The  vortex line with $2\pi$ winding transforms to the domain wall, at which the Green's function has a jump,  $G(\omega=+0,p_x,p_y)- G(\omega=-0,p_x,p_y)\neq 0$. This domain wall terminates on $\pi$ vortices.

\subsection{Flat band near the saddle point}

Here we consider the formation of the Khodel-Shaginyan flat band in the vicinity of a saddle point in the $d=2$ quasiparticle spectrum using the phenomenological Landau theory \cite{Volovik1994} and compare it with the results of numerical simulations of the Hubbard model \cite{Yudin2014}. 

The simple Landau-type functional for the interacting Fermi liquid is
\begin{equation}
{\cal E}\{n({\bf p})\}  = \sum_{\bf p} n_{\bf p} \epsilon^{(0)}_{\bf p} + \frac{1}{2}\sum_{{\bf p},{\bf p}'} f({\bf p},{\bf p}') n_{\bf p}  n_{\bf p}' \,.
\label{LandauTheory}
\end{equation}

\begin{figure}[t]
\centerline{\includegraphics[width=0.7\linewidth]{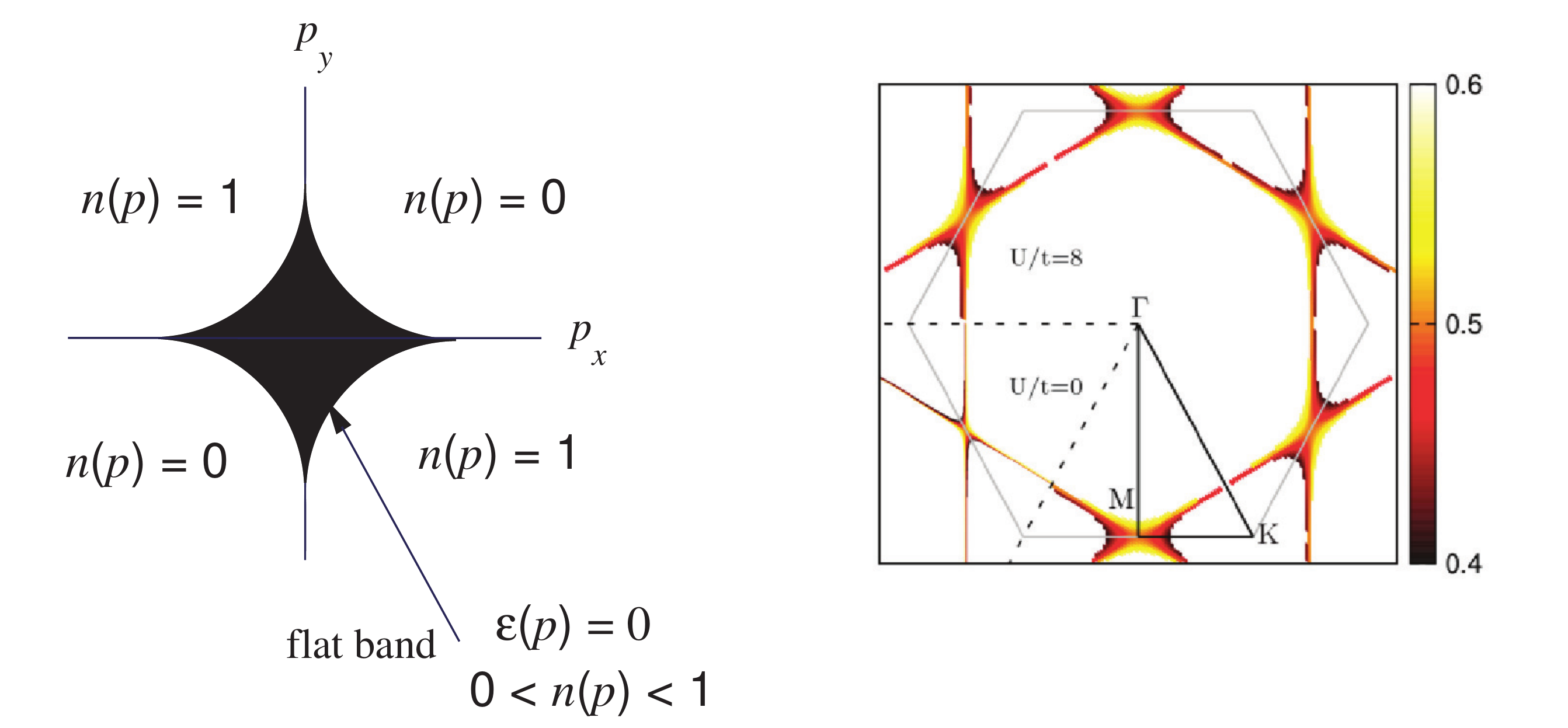}}
\medskip
\caption{Flat band emerging near a saddle point. ({\it Left}): from
  the simplified Landau-type theory in
  Eqs. (\ref{Saddle})-(\ref{VariationSolution2}) with $\mu=0$. The
  flat band is concentrated in the black region.  ({\it Right}): from
  the numerical solution of the Hubbard model \cite{Yudin2014}
  [D. Yudin, {\it et al.}, Phys. Rev. Lett. {\bf 112}, 070403 (2014)],
  showing the spectral function within the reciprocal space of an interacting triangular lattice.
The lower left sextant corresponds to the noninteracting case $U=0$. For large $U$ the band flattening
is clearly seen near the saddle points. }
\label{SaddleFig}
\end{figure}

We illustrate the flat band solution using an even simpler functional with contact interaction:
\begin{equation}
 {\cal E}\{n({\bf p})\} = \sum_{\bf p}\left[\epsilon^{(0)}_{\bf p}  n_{\bf p}  + 
\frac{1}{2} U \left(n_{\bf p}-\frac{1}{2}  \right)^2\right]
\,,
\label{Saddle}
\end{equation}
where $U>0$.
This functional has always a flat band solution with $0< n_{\bf p} <1$:
\begin{equation}
 \epsilon_{\bf p}=\frac{\delta{\cal E}}{\delta n_{\bf p}} =\epsilon^{(0)}_{\bf p}   +U  \left(n_{\bf p}-\frac{1}{2}  \right)=0 \,,
\label{VariationSolution1}
\end{equation}
\begin{equation}
  n_{\bf p}  = \frac{1}{2} - \frac{\epsilon^{(0)}_{\bf p}}{U}~~,~~ 0< n_{\bf p} <1 \,.
\label{VariationSolution2}
\end{equation}
In the vicinity of the saddle point the non-perturbed spectrum (i.e. at $U=0$) has the form $\epsilon^{(0)}_{\bf p}=\frac{p_x p_y}{m} - \mu$. For $\mu \neq 0$, there are two hyperbolic Fermi surfaces. They interconnect at the Lifshitz transition, which   takes place at $\mu=0$. When the interaction $U$ is switched on, the flat bands emerge. Figure \ref{SaddleFig} ({\it left}) demonstrates  the flat band at $\mu=0$.
The same shape of the region with the flat band has been obtained from the numerical simulations of  the Hubbard model \cite{Yudin2014},  see Fig. \ref{SaddleFig} ({\it right}).

\section{Flat bands in topological semimetals}
\label{semimetals}

In the previous section we discuss the fermion condensate -- the flat band, which may emerge due to interactions in the vicinity of a singularity in the non-interacting spectrum.
There are also other ways to generate flat bands or approximate flat
bands. Historically the flat bands first appeared as Landau levels of charged particles in a magnetic field \cite{Landau1930}. Here we discuss the flat bands that
have purely topological origin. They may exist without a magnetic field,  and they are not very sensitive to interactions.
 The flat band may emerge as the surface or interface state in topological semimetals 
\cite{Ryu2002,HeikkilaKopninVolovik2011,SchnyderRyu2011}, which we discuss in this section.
The flat band may also appear at the strained interfaces with misfit
dislocations, which play the role of effective magnetic field. Such
flat bands are discussed  in Sec.~\ref{strained}.

In this section we characterize semimetals that have an internal
spin-like structure. All the examples discussed here can be
characterized via the topological invariant of the form \cite{Volovik2007}  
\begin{equation}
N_1=  {\bf tr} \oint_C \frac{\mathbf dl}{4\pi i} \cdot[\Gamma H^{-1}(\vec{p}) \partial_l H(\vec{p})],
\label{eq:N1}
\end{equation}
where $H$ is the Hamiltonian in the momentum space; $\Gamma$ is a matrix which commutes 
or anti-commutes with the Hamiltonian, such as the
third Pauli matrix  $\sigma_3$ acting on the spin-like degree of freedom; and $C$
is a contour in momentum space, specified separately for each
semimetal.

In semimetals, the flat bands are realized as a particular consequence of the
bulk-boundary correspondence of topological media (see \cite{mong11,essin11,graf13} or
Sec. 22.1 in \cite{volovikbook}). If for example a 3D bulk system
contains Weyl points, then an interface with a topologically trivial material, or with a
material having a different value of topological invariants
contains a line of zeroes -- the Fermi arc \cite{Burkov2011}. The termination points
of the Fermi arc are given by projections of the Weyl points to the interface.
 In the same manner the Dirac lines 
in 3D bulk or Dirac points in 2D bulk give rise to the to nodes of higher dimension at the interface -- 
the flat band \cite{Ryu2002,HeikkilaKopninVolovik2011,SchnyderRyu2011}. The boundaries of the flat band
are determined by the projection of the Dirac line or Dirac points to the interface.
 
\subsection{Topological nodes: Dirac lines and Dirac points}



\begin{figure}[h]
 \begin{center}
  \includegraphics[width=0.3\linewidth]{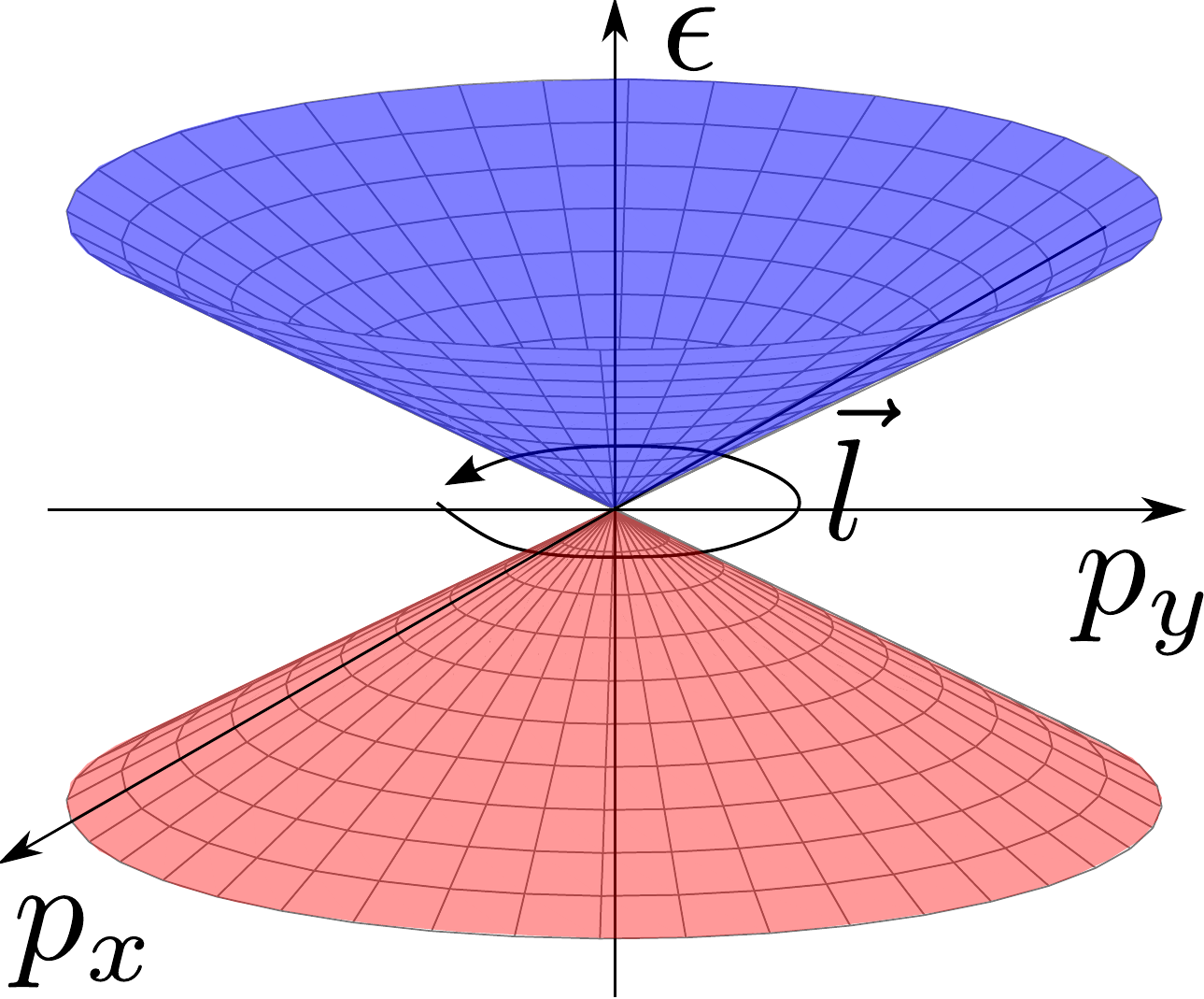}
\end{center}
  \caption{\label{fig:diracpoint}
2D Dirac/Fermi point in momentum space and the line of integration for
the topological invariant $N_1$.
 }
\end{figure}

Let us consider an example semimetal characterized by an even number
of 2-dimensional Dirac points, such as that found in graphene around the two valleys. Close to the Dirac point the Hamiltonian can be written as a 2$\times$2 matrix in (pseudo)spin space,
\begin{equation}
H=v_F \begin{pmatrix} 0 & p_x-i p_y \\ p_x+i p_y & 0\end{pmatrix} = v_F \vec{p} \cdot \vec{\sigma},
\label{eq:Fermipoint}
\end{equation}
where $\vec{p}=(p_x,p_y)$ and $\vec{\sigma}=(\sigma_x,\sigma_y)$,
where $\sigma_j$ are Pauli matrices. The eigenvalues of $H$ satisfy
$\epsilon^2 = v_F^2 |p|^2$ having a node $\epsilon(\vec{p}=0)$ in a
single point in momentum space. To illustrate the topological
protection of this node, let us add a perturbation of the form
$V(\vec{p}) =  \vec{v}(\vec{p}) \cdot \sigma$, where
$\vec{v}=(v_x,v_y)$ does not break the (pseudo)spin symmetry. As a
result, the dispersion becomes $\epsilon^2 = (v_F p_x + v_x)^2+(v_F
p_y+v_y)^2$, which again has a single node at
$(p_x,p_y)=-(v_x,v_y)/v_F$. The only effect of the potential is thus
to shift the node, but not annihilate it. This property can be
expressed via the presence of the topological charge of the form
\eqref{eq:N1}, where $\Gamma=\sigma_z$ and
the contour $C$ goes around the Dirac point in the 2D momentum
space (see Fig.~\ref{fig:diracpoint}). For the Hamiltonian in Eq.~\eqref{eq:Fermipoint}, we get $N_1=1$. In graphene, there are two Dirac points: the first one has the form in Eq.~\eqref{eq:Fermipoint}, and the second is otherwise the same but $p_y \mapsto -p_y$. In that case the second Dirac point has $N_1=-1$. These topological charges stay invariant to perturbations of the form $V(\vec{p})$, as long as we shift the contour of integration along with the shift of the Dirac point, and as long as the two nodes do not merge due to such a perturbation. 

In 3D materials, the topological charge in
Eq.~\eqref{eq:N1}  characterizes lines of nodes -- the Dirac lines. The Dirac lines are readily obtained 
in superfluids and superconductors, where the symmetry operator $\Gamma$ contains the particle-hole symmetry. In particular, the Dirac line exists in the polar phase of superfluid $^3$He \cite{SilaevVolovik2014} 
and  may appear in superconductors without inversion symmetry \cite{SchnyderRyu2011}. 
The topologically stable 
Dirac lines give rise to the topologically protected surface flat band. According to the bulk-surface correspondence,  the boundary 
of the flat band is determined by the projection of the nodal line on to surfaces.
In nonsuperconducting materials the corresponding symmetry which enters Eq.~\eqref{eq:N1} can be only approximate, being violated by spin-orbit interaction, or by the higher order hopping elements. This  leads to formation of approximately flat surface bands as it
happens for example in graphite \cite{heikkila11a}, graphene networks \cite{Weng2014}  and possibly in some other materials
\cite{Kane2015}.

\begin{figure}[h]
 \begin{center}
  \includegraphics[width=0.6\linewidth]{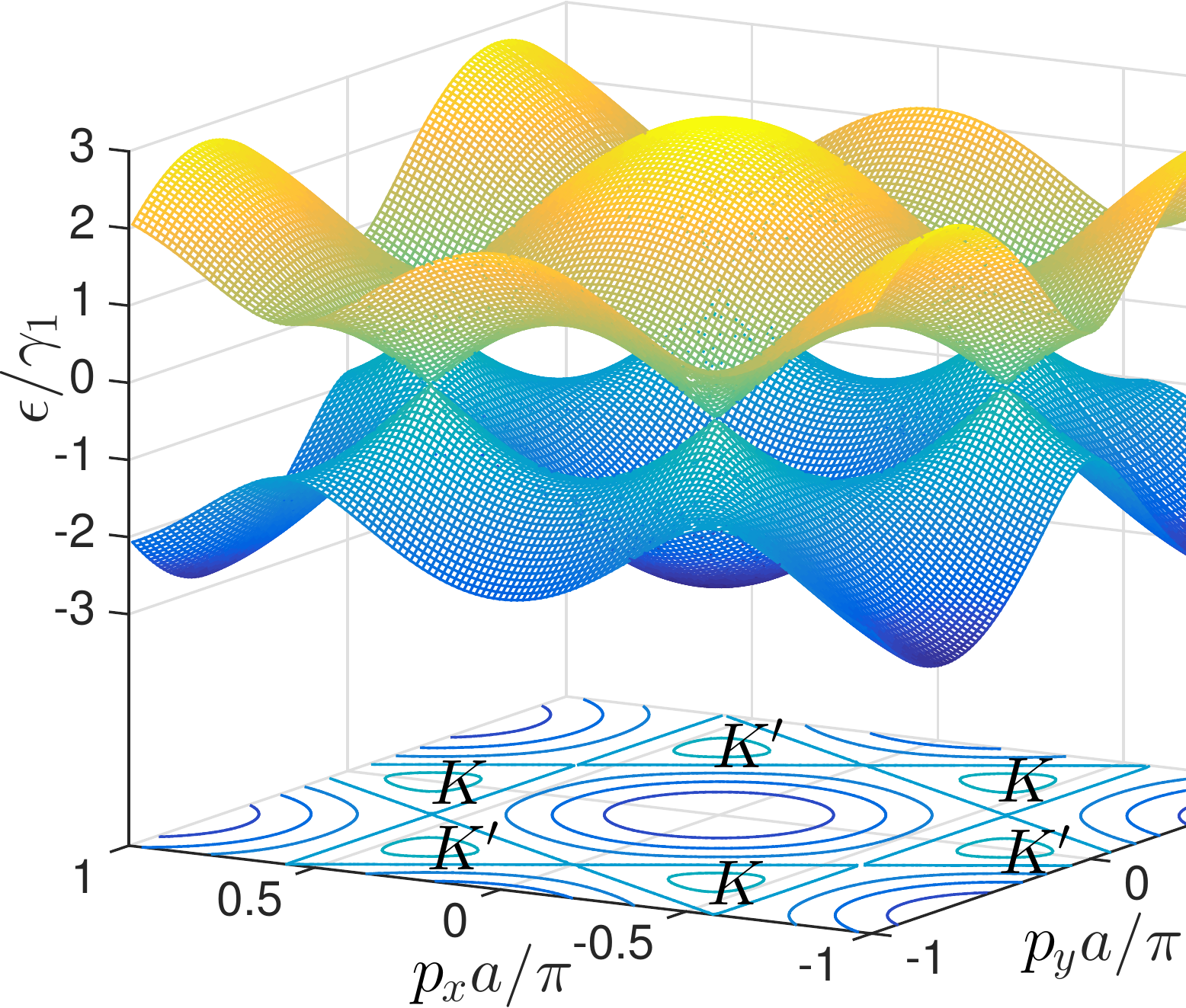}
\end{center}
  \caption{\label{fig:graphenefullspectrum}
Energy spectrum in the conduction and valence bands of graphene. Here
$a$ is the lattice constant, and $\gamma_1$ denotes the
nearest-neighbour hopping parameter in the tight-binding lattice. The
underlying contour plot shows the positions of the Dirac points. Only
two of these points are non-equivalent, the others are connected via
reciprocal lattice vectors.
 }
\end{figure}

We first consider the case of graphene, whose energy spectrum within
the valence and conduction bands is plotted in
Fig.~\ref{fig:graphenefullspectrum}. This spectrum follows for example
from the nearest-neighbour tight-binding model on a honeycomb lattice,
see \cite{Heikkilabook,Katsnelsonbook}. Around the specific points,
marked $K$ and $K'$ in the figure, the low-energy Hamiltonian is of
the form of Eq.~\eqref{eq:Fermipoint}. These points are described by
the topological charge $N_1$, such that $N_1(K)=+1$ and $N_1(K')=-1$. 

\begin{figure}[h]
 \begin{center}
  \includegraphics[width=0.8\linewidth]{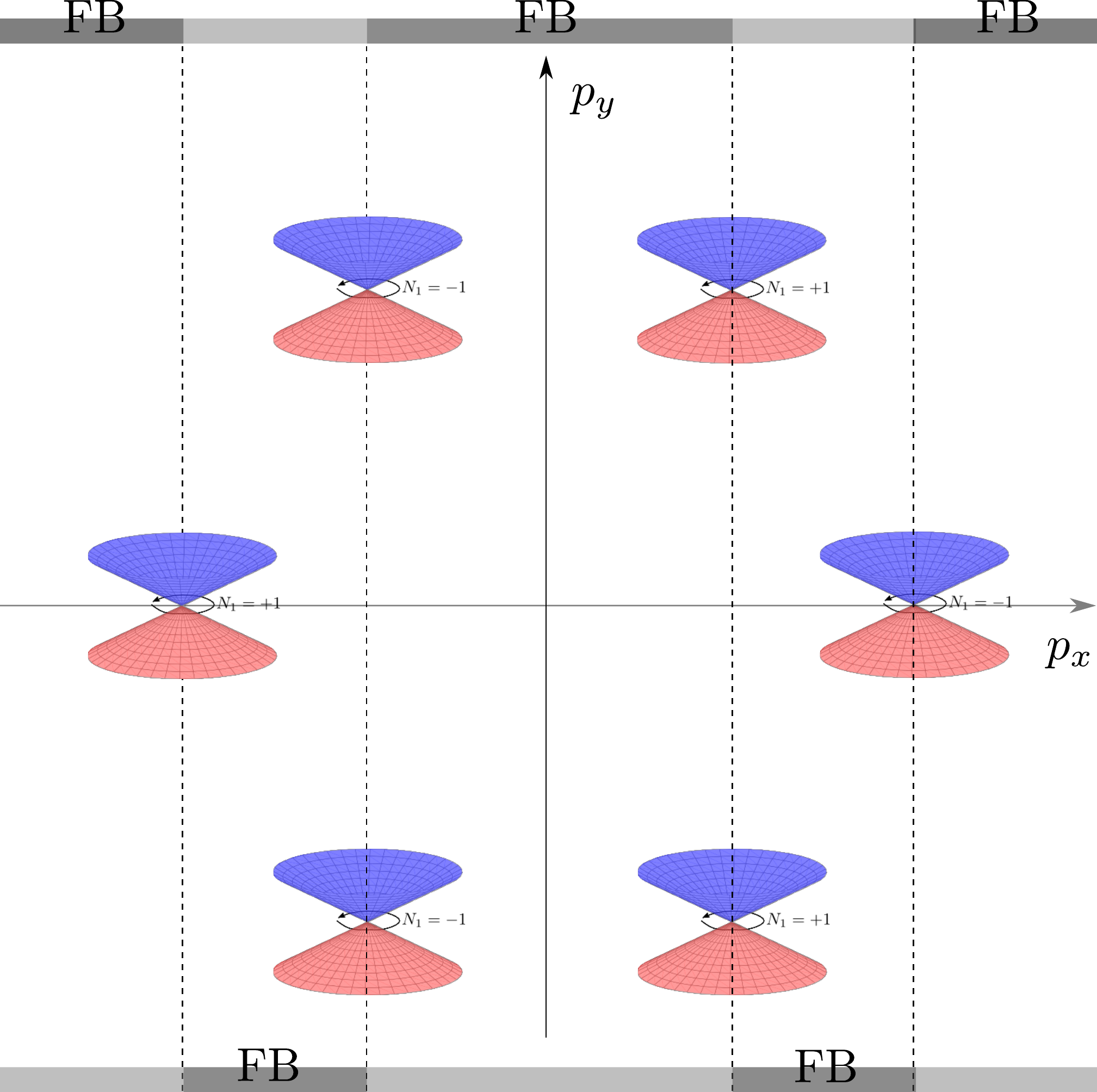}
\end{center}
  \caption{\label{fig:grapheneflatband}
Formation of a flat band in the zigzag (lower part of the figure) or
bearded (upper part) edges of
graphene.}
\end{figure}

Figure
\ref{fig:grapheneflatband} shows the formation of flat bands at the
edges of graphene \cite{Ryu2002}. There, we show the
locations of the $K$ and $K'$ points in the 2d momentum plane, and
consider the presence of an edge placed in the
$x$-direction. Now, the edge marks a boundary between graphene,
which is a nodal semimetal, and vacuum, which is a trivial
insulator. By the bulk-boundary correspondence we may hence expect
flat band states at the edge. Since we maintain translational
invariance along the $x$ direction, $p_x$ remains a good quantum
number and it also parametrizes the edge states and their dispersion $\epsilon_e(p_x)$. 
According to the bulk-boundary correspondence, the projections of the Dirac points
to the boundary determine the end points of the flat band.
From Fig.~\ref{fig:grapheneflatband} it
is clear that the termination points of the flat band must be located at $p_x$
values corresponding to the $p_x$-component of the $K$ and $K'$
points. When $p_x$ crosses these points, the normal to the interface run 
 across either $K$ or $K'$ points and the topological invariant  in
Eq.~\eqref{eq:N1}  along the normal
changes.  However,
from the figure alone one cannot say which of the regions between the
$K$ and $K'$ points contain flat bands and which not. The solution is
to construct the system by repeating a set of infinite chains, and
construct the topological invariant, in this case called the Zak phase
\cite{Delplace2011} for these chains. As shown for example in \cite{Delplace2011,Ryu2002}, the details of this procedure depend on the
microscopic form of the edge, which is either of the ``bearded'' or
the ``zigzag'' type. Figure \ref{fig:grapheneflatband} illustrates the
resulting positions of the flat band dispersion in momentum space. 

If we were to place the interface in the $y$ direction (in which case
we would obtain the ``armchair'' edge), there would be
no flat band due to the fact that the normal to the interface would
run across both $K$ and $K'$ points. In this case the change of the
topological charge across this normal line would be $N_1(K)+N_1(K')=0$.


In the following sections we explain the topological invariants in multilayer graphene structures and discuss the consequent (approximate) flat band surface states. 



\subsection{From graphene to graphite}
In graphite, the coupling between the graphene layers is much weaker than that within the atoms producing the hexagonal lattice in individual layers. Because of this, we may assume that the individual layers remain to be described by the two-dimensional low-energy momentum-space Hamiltonian of the form in Eq.~\eqref{eq:Fermipoint}. In the (most stable) AB stacking, the pair of layers is arranged so that the graphene hexagons are rotated with respect to each other by 30 $^\circ$. As a result, one pair of atoms (say, A on the bottom layer and B on the top layer) reside on top of each other, whereas the other atom of the unit cell (B on the bottom and A on the top) are on the bottom/top of the other hexagon. Therefore, the interlayer coupling between the first pair of atoms is much stronger than the coupling between the second pair. Taking into account only this strongest coupling then produces the Hamiltonian of the form
\begin{equation}
H_K=\begin{pmatrix} v_F \vec{p} \cdot \vec{\sigma} & -\gamma_1 \sigma_\downarrow \\ -\gamma_1^* \sigma_\uparrow & v_F \vec{p} \cdot \vec{\sigma} \end{pmatrix}
\end{equation}
around one of the graphene valleys ($K$-point) for this pair of
layers. Here $\gamma_1$ quantifies the interlayer hopping. Around the
other valley ($K'$ point) the Hamiltonian is
$H_{K'}(p_x,p_y)=H_K(p_x,-p_y)$. It is straightforward to show that
$H_{K/K'}$ have four branches of eigenvectors, two gapped ones (with $\epsilon(|p|=0)=\pm \gamma_1$) and two with a quadratic dispersion around a Fermi node, $\epsilon^4 = |\gamma_1|^2 (p/p_{FB})^4$, where $p_{\rm FB}=|\gamma_1|/v_F$. The corresponding $N_1=2$ ($-2$) for $H_K$ ($H_{K'}$). 

Beyond the bilayer, the graphene layers can be stacked in two qualitatively different ways by respecting the AB stacking for each pair of layers. In Bernal stacking, the line of strongest interlayer coupling is straight, i.e., connects the same atoms in each layer whereas for rhombohedral (ABC) stacking, it follows an armchair-type pattern, i.e., the strongly coupled atoms differ between neighbouring pairs of layers. The corresponding elements of the momentum space Hamiltonians coupling layers $n$ and $m$ for 2D momenta around the $K$ point are 
\begin{align}
H^{\rm Bernal}_{mn}&=v_F \vec{p} \cdot \vec{\sigma} \delta_{mn} - \gamma_1 \delta_{m,n+1} \left\{\sigma_\uparrow [(-1)^m-1]+\sigma_\downarrow [(-1)^m+1]\right\}/2+{\rm h.c.}\\
H^{\rm rhombohedral}_{mn}&=v_F \vec{p} \cdot \vec{\sigma} \delta_{mn} - \gamma_1 \delta_{m,n+1} \sigma_\uparrow +{\rm h.c.}
\end{align}
In $H^{\rm Bernal}$ the terms in square brackets take care of indexing the even and odd layers separately. Note that since the choice of the $A/B$ indices for the graphene sublattice atoms is arbitrary, we could have also chosen to write the coupling in $H^{\rm rhombohedral}$ in terms of $\sigma_\uparrow$ instead of $\sigma_\downarrow$. Formally, these two choices correspond to different signs of the bulk topological invariants (see below), but they have a meaning only in the presence of (Bernal) stacking faults that change the coupling around some particular interface.

\subsection{Bulk Fermi line in Bernal graphite}

\begin{figure}[h]
 \begin{center}
  \includegraphics[width=0.7\linewidth]{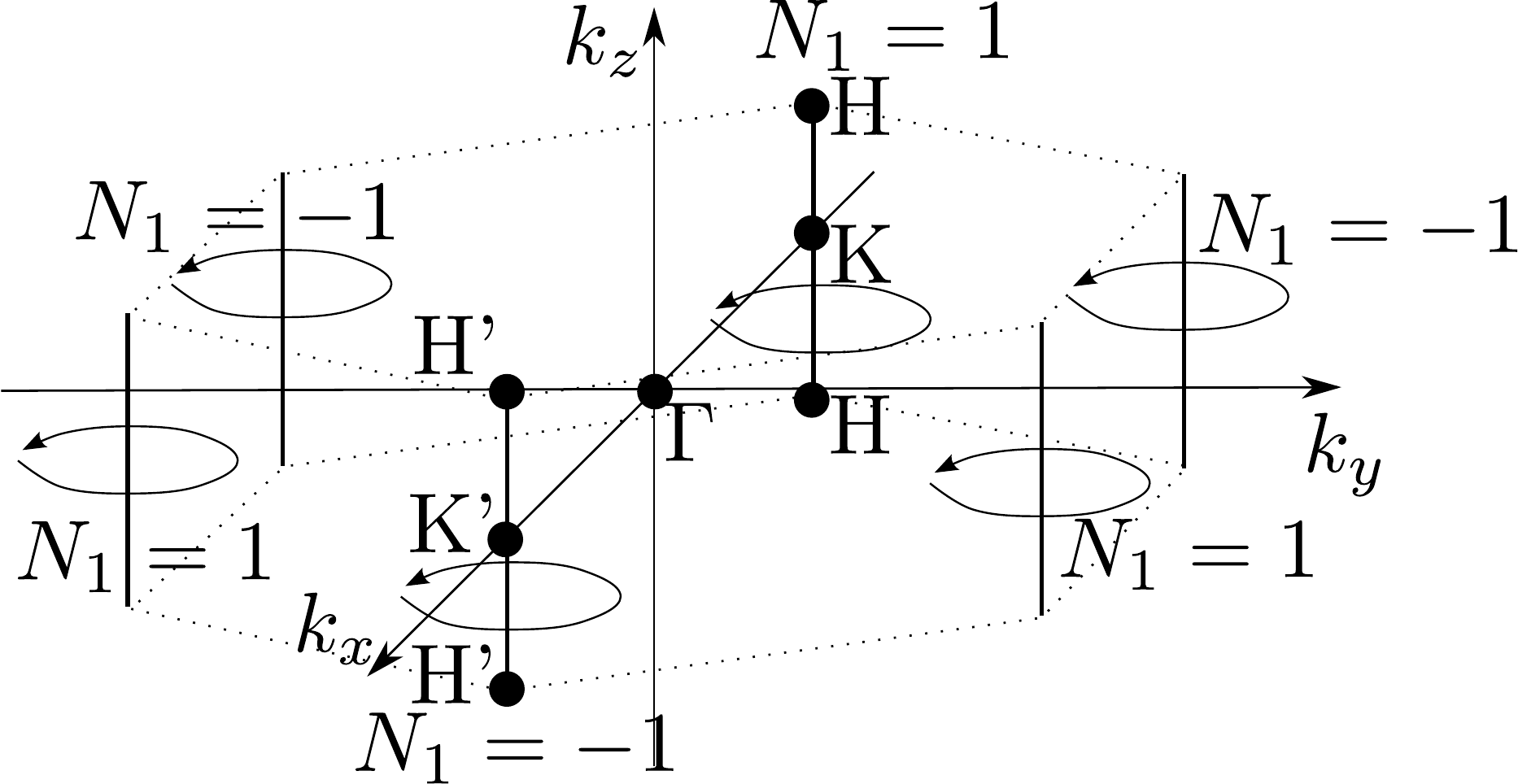}
\end{center}
\label{fig:bernalgraphitefermilines}
  \caption{
Fermi lines in Bernal graphite within the nearest-neighbour interlayer
approximation. This picture expands the 2D momentum space
($p_x/p_y$ plane) of graphene to the 3D space where $p_z$ runs between $-\hbar\pi/a$ to
$\hbar\pi/a$ ($a$ is the interlayer spacing). 
 }
\end{figure}

We first derive the bulk dispersion of Bernal graphite by including only the strongest interlayer hopping term $\gamma_1$. 
We make the bulk Ansatz
\begin{equation}
\psi_n^T=\begin{cases} \begin{pmatrix} \alpha_o & \beta_o \end{pmatrix} e^{ip_z na/\hbar},\quad n \text{ odd}\\
\begin{pmatrix} \alpha_e & \beta_e \end{pmatrix}, \quad n \text{ even} \end{cases}
\end{equation}
for the wave function on the $n$-th layer. Here $p_z$ is the momentum in the direction perpendicular to the layers and $a$ is the distance between the layers. This is an eigenfunction with energy $\epsilon$ provided that the coefficients $\alpha_{e/o}$, $\beta_{e/o}$ satisfy
\begin{equation}
\label{eq:Hb}
\underbrace{
\begin{pmatrix} 0 & v_F p_- & 0 & -2 \gamma_1 \cos(p_z a/\hbar)\\
v_F p_+ & 0 & 0 & 0\\
0 & 0 & 0 & v_F p_-\\
-2\gamma_1^* \cos(p_z a/\hbar) & 0 & v_F p_+ & 0
\end{pmatrix}}_{H_B}
\begin{pmatrix} \alpha_o \\ \beta_o \\ \alpha_e \\ \beta_e \end{pmatrix} = \epsilon \begin{pmatrix} \alpha_o \\ \beta_o \\ \alpha_e \\ \beta_e \end{pmatrix},
\end{equation}
where $p_{\pm} = p_x \pm i p_y \equiv p_\perp e^{i\phi}$,
$p_{\perp}\ge 0$. This has (four) zero-energy solutions
at $p_x=p_y=0$, regardless of the value of $p_z$. Within this
approximation, Bernal graphite has therefore two Dirac lines, running
through the $K$ and $K'$ points of the 2D graphene band structure, or
between the H points in the 3D graphite band structure, see
Fig.~\ref{fig:bernalgraphitefermilines}. Including higher-order
hopping terms in the Hamiltonian expands this line into electron and
hole pockets \cite{mcclure57}. However, let us first analyze the topology of $H_B$. The
topological charge in this case is of the form of Eq.~\eqref{eq:N1},
where the contour $C$ runs around the Dirac lines as indicated in
Fig.~8, and $\sigma_z$ should be
replaced by $(\sigma_z \otimes 1)$.  This produces
  $N_{1}=\pm 2$, where the factor 2 takes care of
the additional layer degree of freedom in Eq.~\eqref{eq:Hb}.

This is thus nothing but the generalization of the graphene
topological charge to graphite and it yields $N_1=\pm 2$ for the
momenta along the lines $H-K-H/H'-K'-H$ \cite{mcclure57}, where the
$H^{(\prime)}$ point is shifted from the $K^{(\prime)}$ point in the
$p_z$ direction by $\hbar\pi/a$. Based on the
bulk-boundary correspondence we may therefore expect to have surface
states at the lateral boundaries of Bernal graphite, as extensions of
the flat band states in zigzag graphene, but now the flat band extends
throughout the first Brillouin zone in the $p_z$ direction.

\begin{figure}[t]
\centerline{\includegraphics[width=0.9\linewidth]{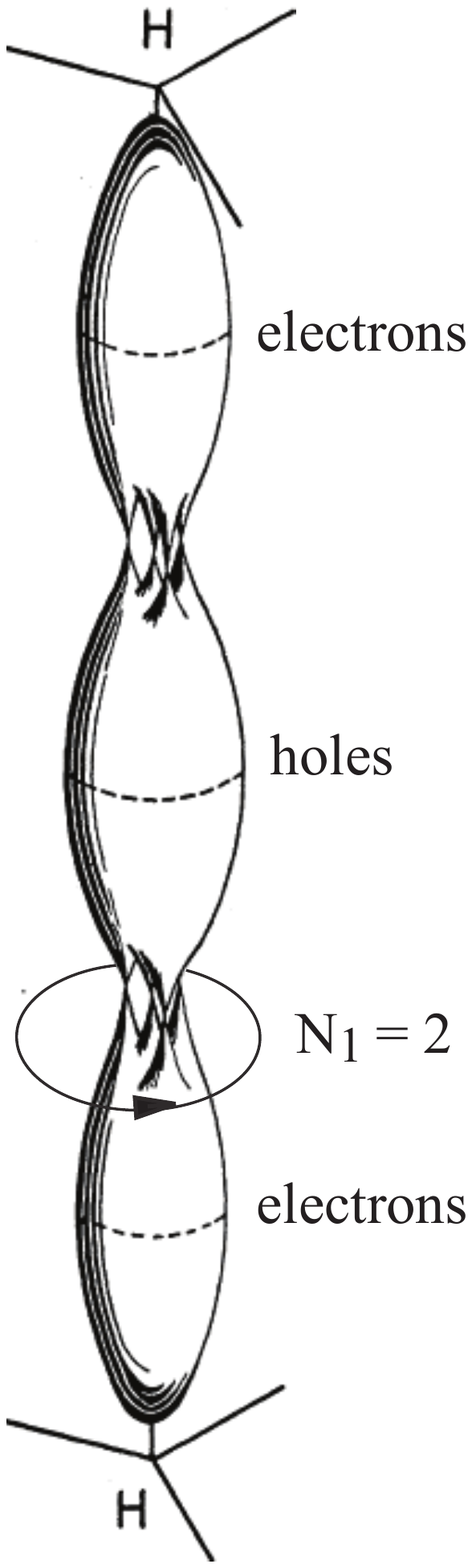}}
\medskip
\caption{Graphite Fermi surface from Ref. \cite{mcclure57}.
It consists of the electron and hole pockets connected at four points. Each point is described by 
topological invariant in Eq.~\eqref{eq:N1}: the point in the center 
has $N_1=-1$ and the other three points have $N_1=1$. Such structure originates from the nodal line 
with $N_1=2$ in the Hamiltonian (\ref{eq:Hb}), when the higher order hopping elements are taken into account.
}
\label{GraphiteFSFig}
\end{figure}

Let us now consider the effect of higher-order hopping elements. 
First they split the Dirac line with multiple charge $N_1=2$ into elementary Dirac lines. The natural splitting would be into two  lines with  charges $N_1=1$. However, the situation is more interesting, see Fig. \ref{GraphiteFSFig} and Refs. \cite{mcclure57,Mikitik2008}. The $N_1=2$ line splits into 4 Dirac lines: three lines with trigonal arrangement have $N_1=1$, while the central line has $N_1=-1$. The total topological invariant remains the same, $N_1=1+1+1-1=2$. 
In addition, the higher-order hoppings break the symmetry $\Gamma$ \cite{heikkila11a} underlying the topological protection. Without the topological protection
the four Dirac lines are expanded into  four pairs of electron and hole pockets, which  touch each other at four points. As a result the surface states of Bernal graphite cease to be flat on the energy scale related to those higher-order hoppings. 


\begin{figure}[h]
 \begin{center}
  \includegraphics[width=0.7\linewidth]{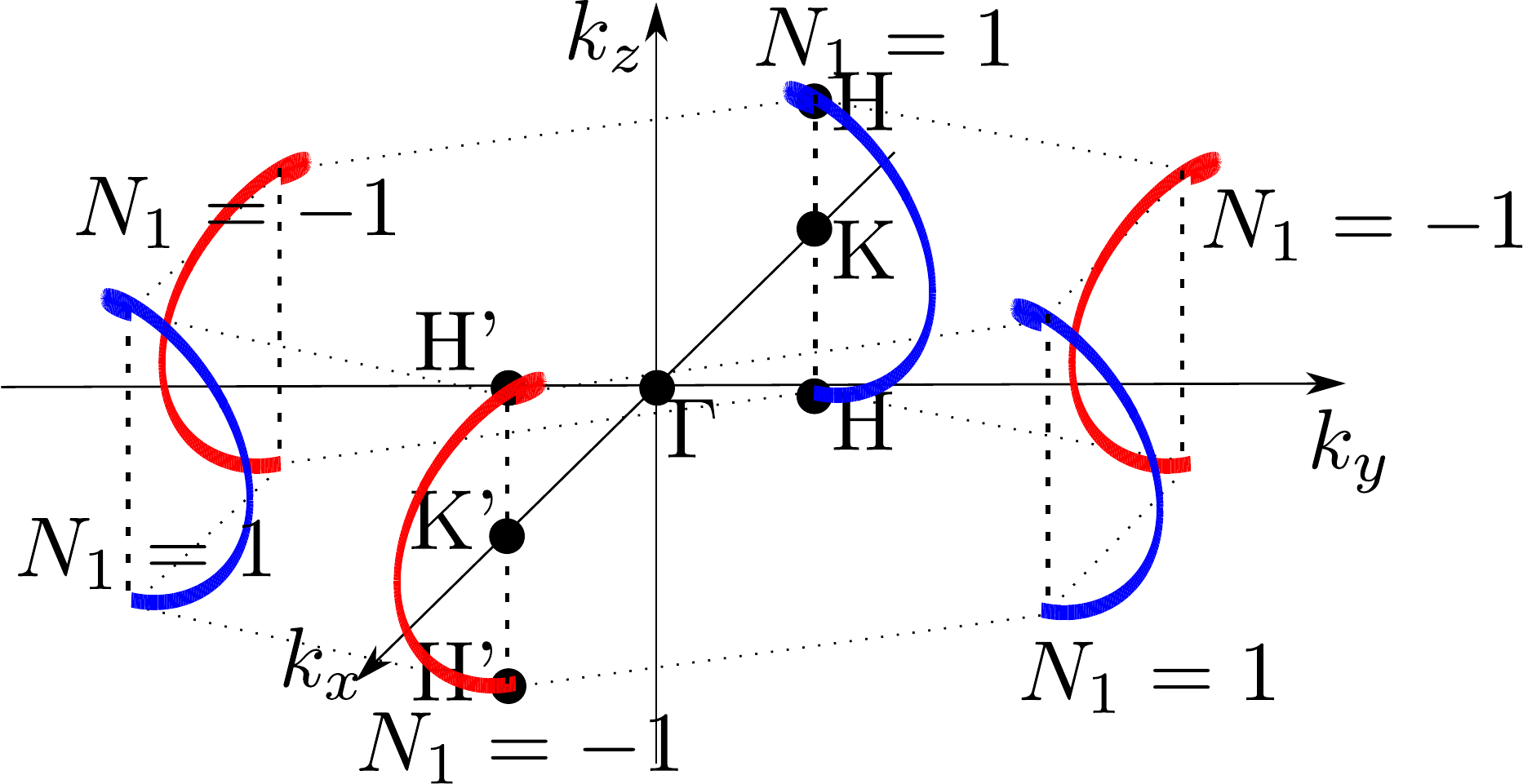}
\end{center}
  \caption{\label{fig:rhgfermilines}
Fermi spirals in rhombohedral graphite.
 }
\end{figure}

\subsection{Spiral Fermi line in rhombohedral graphite}
To obtain the bulk dispersion of rhombohedral graphite, it is enough to make the Ansatz
\begin{equation}
\psi_n=\begin{pmatrix} \alpha \\\beta \end{pmatrix} e^{i p_z n a/\hbar}.
\end{equation}
This is an eigenfunction of $H^{\rm rhombohedral}$ if $\alpha$ and $\beta$ satisfy
\begin{equation}
\underbrace{\begin{pmatrix} 0 & v_Fp_- - \gamma_1 \exp(i p_z a/\hbar)
    \\ v_F p_+ - \gamma_1^* \exp(-i p_z a/\hbar) & 0 \end{pmatrix}}_{H_{\rm RHG}} \begin{pmatrix} \alpha \\ \beta \end{pmatrix} = \epsilon \begin{pmatrix} \alpha \\ \beta \end{pmatrix}
\end{equation}
for some energy $\epsilon$. The zero-energy solutions are found along a spiral in the momentum space,
\begin{equation}
p_z a = \arctan\left(\frac{p_y}{p_x}\right)=\phi, \quad v_F p_\perp=|\gamma_1|.
\label{eq:spiral}
\end{equation}
In this case the topological charge equals the spiral helicity, and
can be defined by Eq.~\eqref{eq:N1} where the integration contour runs
over the 1st Brillouin zone ($-\pi/a$ to $\pi/a$) in the
$p_z$-direction. 
For $p_z$ running close to the line from the H point via K back to H,
$N_1=1$ whenever the transverse momentum $p_x \pm i p_y$ is inside the spiral, i.e., $p_\perp<|\gamma_1|/v_F$. For larger momenta away from the $H-K-H$ line, $N_1=0$. On the other hand, for the path $H'-K'-H'$ the spiral is defined by Eq.~\eqref{eq:spiral}, where $p_y \leftrightarrow -p_y$, inverting the helicity and the sign of $N_1$ to -1.

As a result of the existence of the non-trivial topological charge,
the surface layers of rhombohedral graphite contain flat bands, i.e.,
$\epsilon(p_\perp < |\gamma_1|/v_F)=0$. This can be understood as follows.
For each ${\bf p}_\perp$ with $p_\perp \neq |\gamma_1|/v_F$, the Hamiltonian $H_{{\bf p}_\perp}(p_z)$
describes a 1D  insulator. For $p_\perp < |\gamma_1|/v_F$, this insulator is topological, since
it has the topological charge $N_1({\bf p}_\perp)=1$. According to the bulk-boundary correspondence, each topological insulator has an edge state with zero energy. These states form a flat band in the region $p_\perp < |\gamma_1|/v_F$.

\section{Flat band at strained interfaces}
\label{strained}

Recently another possible source of the topological flat band has been discussed in two materials:
 highly oriented pyrolytic Bernal  graphite (HOPG) \cite{EsquinaziHeikkila2014} and heterostructures SnTe/PbTe, PbTe/PbS, PbTe/PbSe, and PbTe/YbS consisting of a topological crystalline insulator and a trivial insulator \cite{TangFu2014}. In both cases the flat band
comes from a misfit dislocation array, which is spontaneously formed at the interface between two  crystals due to the lattice mismatch.  In Ref. \cite{EsquinaziHeikkila2014} the lattice of screw dislocations has been considered, which emerges at the interface between two domains of HOPG with different orientations of crystal axes. In Ref. \cite{TangFu2014} the misfit dislocation array is formed at the interface between topological and trivial insulators. 
\begin{figure}[t]
\centerline{\includegraphics[width=0.7\linewidth]{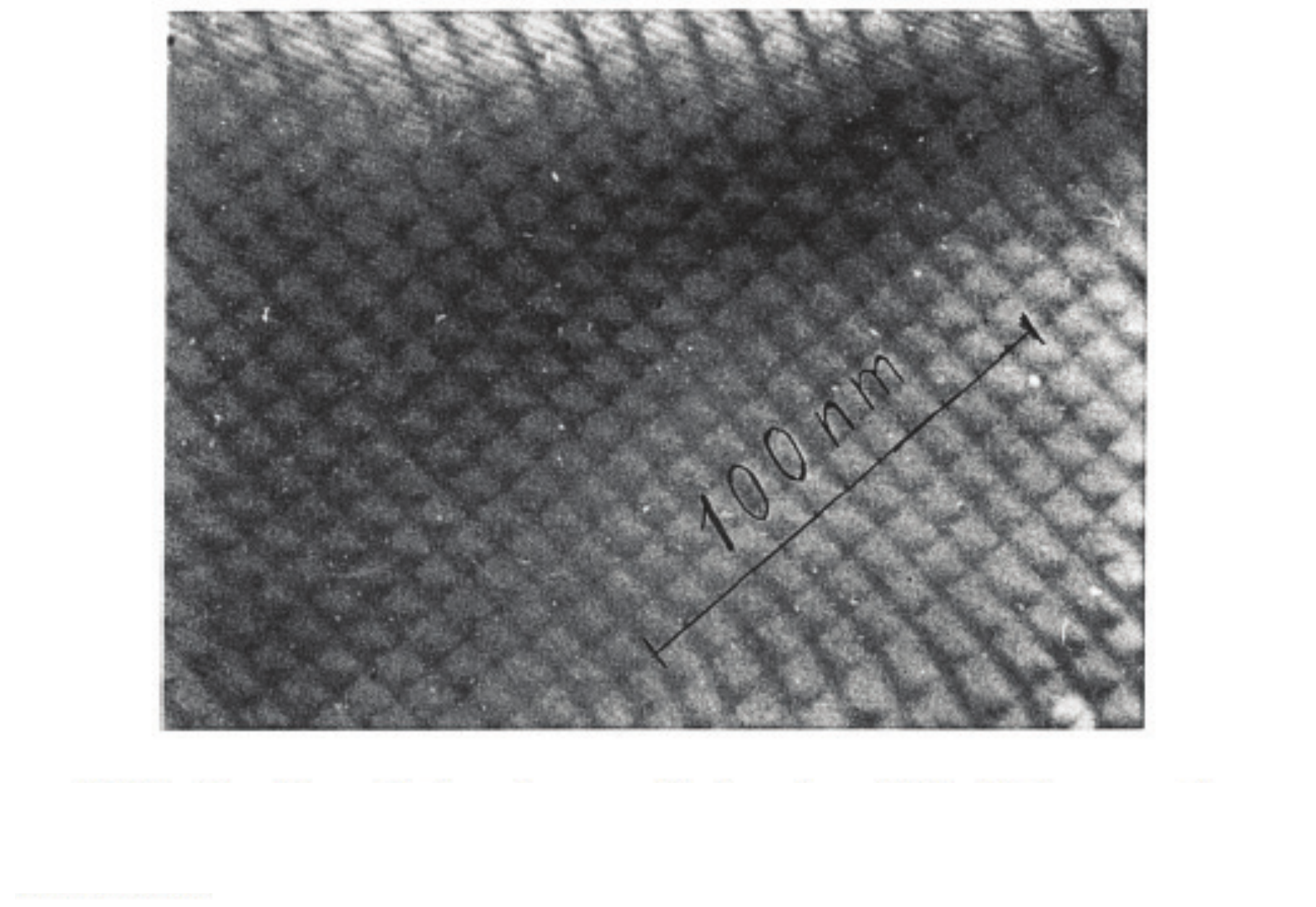}}
\medskip
\caption{
Misfit dislocation grid at the interface from [N. Ya. Fogel,
Phys. Rev. B {\bf 66}, 174513 (2002)] \cite{Fogel2002}.
}
\label{DislocationGridFig}
\end{figure}

The above two systems exhibit similar phenomenon. In both cases 
superconductivity related with the interfaces has been found \cite{Fogel2002,esquinazireferences}. The reported transition temperature essentially exceeds  the typical transition temperature expected for the bulk materials. A possible origin of this phenomenon is the flat band at the interfaces, where the transition temperature could be proportional to the coupling constant and the area of the flat band.

The topological origin of the flat bands in these systems can be understood either  in terms of the overlapping of the 1D flat bands formed within the dislocations or using the following consideration. 
In case of the heterostructures, on one side of the interface the
insulator is topological, and thus the interface contains Dirac
fermions. The strain at the interface (with its dislocations) acts on
Dirac fermions as the effective magnetic field. Such emergent field is
now extensively discussed for strained graphene, see e.g. the most
recent paper Ref. \cite{VolovikZubkov2014} and references therein. The
effective magnetic field $B$ produces the required  flat band, since
the first Landau level for massless Dirac fermions has zero
energy. When the period decreases, the field  $B$ increases, giving
rise to enhanced density of states, which is proportional to $B$. This
is  equivalent to increase of the area of flat band in the scenario
discussed in Sec. \ref{semimetals}. Such increase of transition
temperature is discussed in Ref. \cite{TangFu2014}. 

\begin{figure}[t]
\centerline{\includegraphics[width=0.7\linewidth]{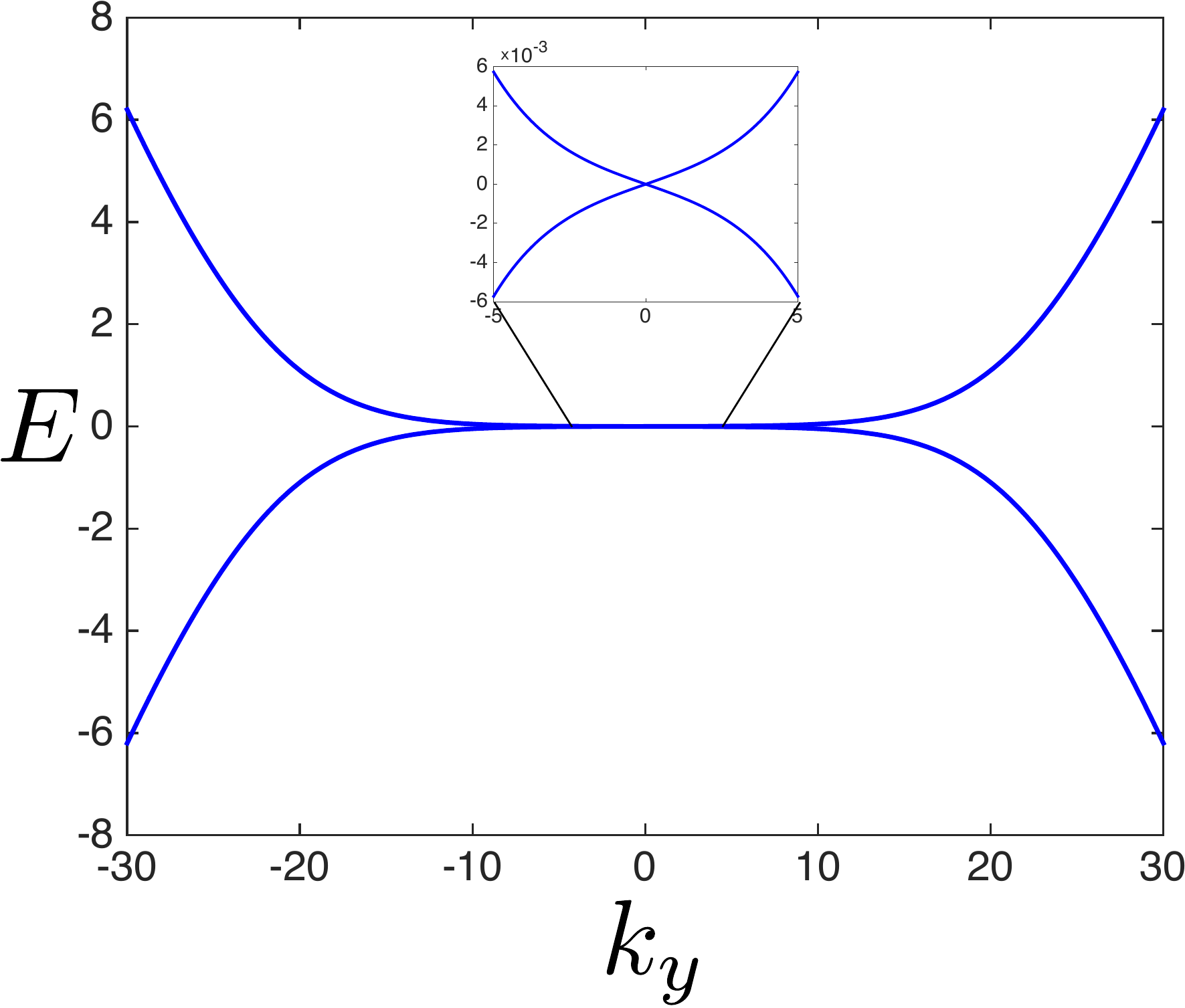}}
\medskip
\caption{
Spectrum of Eq.~\eqref{eq:tangfu} with $\beta=30$. Inset shows the
(approximate) linear
dispersion with low values of $k_y$, and with a speed as in Eq.~\eqref{eq:slope}.
}
\label{fig:tangfuspectrum}
\end{figure}

In particular, Ref.~\cite{TangFu2014} considered the case where the
massless Dirac Hamiltonian of the topological insulator experience a
periodic field. They arrived at the Hamiltonian of the form
\begin{equation}
H_{TF}=-i \partial_x \sigma_x + [k_y- A(x)] \sigma_y,
\label{eq:tangfu}
\end{equation}
written in terms of (scaled) momentum $k_y$ along the dislocation, and
a scaled strain-induced gauge field $A(x)$ satisfying $A(x+d)=A(x)$
and $\int_0^d A(x) dx=0$. In the case of
Ref.~\cite{TangFu2014}, $A(x)=\beta \cos(2\pi x/d)$, but the approach
works for a more general periodic vector potential as well. Due to the position dependent
vector potential, there is no translation symmetry in the $x$
direction and the momentum in the $x$-direction is not a good quantum
number. However, due to the periodicity of the potential we can use
Bloch's theorem and define the pseudomomentum $\tilde k_x$. This
allows for calculating the spectrum of Eq.~\eqref{eq:tangfu}, plotted
in Fig.~\ref{fig:tangfuspectrum} and exhibiting (an approximate) flat band for $k_y <
\beta/2$ and for all values of $k_x \in [-\pi/d,\pi/d]$. 

The emergence of the approximate flat band in Eq.~\eqref{eq:tangfu} can be
understood by first considering the case $k_y=0$. In that case the
Hamiltonian has two (unnormalized) zero-energy solutions,
\begin{align*}
\psi_+&=\begin{pmatrix}0 \\ 1 \end{pmatrix} \exp[\int_0^x A(x') dx']\\
\psi_-&=\begin{pmatrix}1 \\ 0 \end{pmatrix} \exp[-\int_0^x A(x') dx'].
\end{align*}
Let us include the term $H_1= k_y \sigma_y$ as a perturbation. The second order secular
equation produces the $2 \times 2$ Hamiltonian constructed from the
matrix elements of the Hamiltonian $H_1$:
$$
\begin{pmatrix} 
0 & H_{+-}^{(1)}\\
H_{-+}^{(1)} & 0
\end{pmatrix}
= c \begin{pmatrix}0 & k_y\\k_y & 0\end{pmatrix}
$$
with the spectrum $E=\pm c k_y$. The slope $c$ of the spectrum is
obtained from the matrix elements of the normalized eigenfunctions
\begin{equation*}
c=\frac{\int_{-\infty}^{\infty} dx 1}{\int_{-\infty}^\infty dx
  \exp[2\int_0^x A(x') dx']} = \frac{d}{\int_0^d e^{2\int_0^x A(x')
    dx'}}.
\label{eq:slope}
\end{equation*}
For $A(x)=\beta \cos(x)$ we thus get $c=1/I_0(2\beta)$, where $I_0(x)$
is the zeroth Bessell function of the first kind. When $c \ll 1$,
requiring $\beta \gg 1$, the result is an approximate flat band.

Alternatively, we may consider the case of a periodic line
dislocation. There, the vector potential is proportional to the strain
field, $A(x)=\alpha_1 u_{xx}(x)+\alpha_2 u_{yy}(x)$, where $\alpha_1$
and $\alpha_2$ are coupling constants and
\begin{align*}
u_{xx}(x)&=\frac{bz}{2\pi (1-\nu)}\frac{3x^2+z^2}{(x^2+z^2)^2}\\
u_{yy}(x)&=\frac{bz \nu}{\pi(1-\nu)} \frac{1}{x^2+z^2},
\end{align*}
where $b$ is the size of the Burgers vector of the dislocation, $z$ is
the distance from the dislocation plane, and $\nu$ is the Poisson
ratio. We also assume that such dislocations repeat after each $d$. In
this case we may estimate the slope $c$ for $d \gg z$ by
\begin{equation*}
c \approx \exp\left[-2 \int_0^\infty A(x)\right] = \exp\left[-\frac{b (\alpha_1+\nu
  \alpha_2)}{1-\nu}\right].
\end{equation*}
The flat band dispersion where $c \ll 1$ thus requires either $\nu
\approx 1$ or coupling constants $\alpha_i \gg 1/b$. Note that this
result show that the estimates done in Ref.~\cite{TangFu2014} are
overoptimistic, as there $\alpha b \approx 1$. 

\section{Superconductivity in graphene or graphite?}
As discussed in Sec.~\ref{semimetals}, bulk neutral graphite and graphene are typically
considered to have a very low density of states. This is why the
occurrence of superconductivity in these bulk systems would require a strong doping,
shifting the Fermi energy to a finite value and thereby increasing the 
density of states. Let us consider the example case of a 2D Fermi line
with the superconducting coupling $g$, but with a shifted Fermi level
by a value $\mu$. In this case Eq.~\eqref{selfconsistent2} yields for $\mu=0$
\cite{kopnin08}
\begin{equation}
\Delta =(a \epsilon_{uv}-1/a)/2 > 0,
\end{equation}
where $a=g/(\pi \hbar^2 v_F^2)$. This is non-zero only for a large
coupling strength $a > 1/\epsilon_{uv}$. For weak coupling
$|a|<1/\xi_c$, superconductivity is enhanced by doping, because for
non-zero $\mu$ the solution is $\Delta =
2|\mu|\exp(-\frac{1/a-\xi_c}{|\mu|}-1)$. 

This strategy has been followed by some graphite experimentalists. For
example Ref.~\cite{weller05} demonstrates bulk superconductivity of two graphite
intercalation compounds C$_6$Yb and C$_6$Ca with critical temperatures
of 6.5 and 11.5 K, respectively. There one of the effects from the intercalant
layers is the charge transfer, i.e., doping the graphite. 

The situation is opposite in flat band systems: doping does not aid
superconductivity, but rather decreases the critical temperature. Fith
a non-zero chemical potential $\mu$, the gap becomes
\begin{equation}
\Delta=\sqrt{\Delta_{\rm FB}^2-\mu^2},
\end{equation}
where $\Delta_{\rm FB}$ is the gap obtained for $\mu=0$, as in
Eq.~\eqref{linear}. 

There are experimental indications for the presence of superconductivity
at graphite interfaces \cite{esquinazireferences}. In this case, the
interfaces form between differently oriented regions of Bernal
graphite, so that the graphite $c$-axis ($p_z$ in our work here) runs
perpendicular to these surfaces. These observations hence cannot be
explained by the flat bands at the lateral interfaces (along $p_x$ or $p_y$) as one would
expect in Bernal graphite. There are also no direct indications of the
presence of rhombohedral stacking in these systems, but the
resolution of the imaging does not allow ruling out some rhombohedral
regions close to the interfaces exhibiting superconducting properties. Note that rhombohedral stacking has been found in highly oriented pyrolytic
graphite, see for example \cite{Lin2012,Hattendorf2013}. The remaining scenario is
based on the formation of an array of dislocations (Sec.~\ref{strained}) at these surfaces
as the surface layers try to adapt to the neighbouring graphite
planes. 

The discussion in Sec.~\ref{pairing} concerns only the relation of the
mean field order parameter with normal-state electronic spectrum. The
details of the flat band superconducting state, such as the
quasiparticle spectrum, supercurrent and collective modes, depend 
on the way the flat band is formed. The case of rhombohedral graphite
has been worked out in Refs.~\cite{kopnin11,kopnin11b,kopnin13} and
summarized in \cite{kopninheikkila14}. The density of states in the
superconducting state was computed in \cite{Munoz2013}. In that case the
superconducting spectrum is not flat, but rather has an inverse
parabolic shape and exhibits a minigap whose value is inversely
proportional to the distance between adjacent surfaces. The
supercurrent is characterized by a large critical current that is linearly 
proportional to $\Delta$. However, in this model the supercurrent does
not only flow along the surfaces, but also between them. On the other
hand it is clear that not all types of flat bands can describe a state with a
non-vanishing critical current. For example, an entirely isolated
region in momentum space having a flat band describes a set of
completely localized electrons, which cannot be made to carry
supercurrent. 

In this chapter we have concentrated on illustrating the results for
mean-field models of superconductivity on flat bands. As discussed in
\cite{kopninheikkila14}, fluctuation effects are much more important
in flat band superconductors than in ordinary BCS superconductors. The
nature of the fluctuation response depends on the mechanism of the
flat band formation, and the ensuing superconducting spectrum. The
details of such fluctuation effects we leave for further work.

As we have discussed in this chapter, there are many ways of realizing
(approximate) flat bands. Some of these realizations may be relevant
in interfaces of graphite, where indications of superconductivity with
a very high critical temperature have been observed. It remains to be
demonstrated whether these properties can be explained within the flat
band scenario of superconductivity.

We acknowledge Pablo Esquinazi, Ville Kauppila and Timo Hyart for
discussions. This work was supported by the Academy of Finland through
its Center of Excellence program, and by the European Research Council
(Grant No. 240362-Heattronics).

\end{document}